\documentclass[]{aa}
\usepackage{epsfig}
\usepackage{natbib}
\usepackage{txfonts}

\newcommand{\water}{$\mbox{H}_{2}\mbox{O}$}

\newcommand{\kms}{$\mbox{km~s}^{-1}$}

\newcommand{\msol}{\mbox{M\hbox{$_\odot$}}}

\newcommand{\arcs}{$^{\prime \prime}$}

\newcommand{\etal}{{et~al.}}
\begin{document}

\title{First VLBI observations of methanol maser polarisation, 
  in G339.88-1.2}

\author{Dodson, R. \inst{1}}

\institute{Observatorio Astron\'omico Nacional,\\ Apartado 112, 
Alcala de Henares, 28803, Spain\\ \email{r.dodson@oan.es}}

\abstract
{}
{We
  investigate class II methanol masers and the environment in which they
  form with the Long
  Baseline Array (LBA).}
{Using full polarisation VLBI, we're able to measure the magnetic
 field directions so as to distinguish between the two main models of the
 environment in which methanol masers form: disks or shocks.}
{We present polarised images of the methanol maser source G339.88-1.2,
made with the LBA at 6.7-GHz.
 With these first polarisation maps made with the LBA, which
 successfully reproduce observations with the ATCA confirming the  
 new AIPS code, a new technique for Southern VLBI is opened. The 
 magnetic field directions found are inconstant with
 methanol masers arising in disks for the majority of the emission.}
{}

\maketitle

\section{Introduction}
\label{s:intro}

Massive stars play a crucial role in the evolution of galaxies, but
the processes leading to massive star formation are poorly understood,
e.g. \citet{garay_99}, \citet{mckee_03}. One of the major questions is whether
or not they form from the accretion or via congregation of previously
formed smaller bodies. Masers, being compact, bright, and at exact
frequencies, are sensitive to the local physical
environment, making them ideal probes of the areas in and
around massive star-forming regions.
Interstellar masers of three molecules, \water , OH and methanol, are
both powerful and widely associated with active high-mass star
formation regions \citep{forster_89,menten_91}. However, both \water\
and OH masers are quite common and also found towards other objects
such as low mass star-forming regions \citep{furuya_01}, evolved
stars \citep{wilson_72} and the centres of galaxies
\citep{santos_79}, unlike methanol masers which are exclusively linked
to high mass star formation \citep{minier_03}.

The two strongest class II methanol masers (i.e. those closely
associated with OH masers and strong far-infrared emission) were
discovered at 12.179~GHz by \cite{batrla_87} and at 6.669~GHz by
\cite{menten_91}. These methanol masers are closely associated with
high-mass star forming regions \citep{norris_88, walsh_97}.
The brightness of methanol masers allows high resolution
interferometric observations to be made, which provide accurate
measurements of position, velocity and dimensions of the individual
maser components. Previous interferometric observations have generally
used connected element interferometers \citep{norris_93, phillips_98,
walsh_98} that offer resolutions of the order of arcseconds (which can
be improved upon by super-resolving the velocity features, see
\cite{phillips_98}), though a number of VLBI (Very Long Baseline
Interferometry) experiments (e.g. \cite{menten_88, menten_92,
norris_98, minier_00, moscadelli_99, dodson_04}), which provide
milliarcsecond resolution, have been performed.  The VLBA covers
12.2-GHz, while the Australian Long Baseline Array (LBA) and the
European VLBI Network (EVN) cover the 6.7-GHz transition. Furthermore
high resolution, connected-element interferometry, imaging has been
done with MERLIN \citep{vlemmings_06} (6.7-GHz) and the
Parkes-Tidbinbilla Interferometer \citep{norris_88} (12.2-GHz). One of
the early successes of methanol maser imaging was to provide strong
support for the accretion model. It was claimed from such observations
that methanol masers formed predominately linear structures, and these
marked the accretion disks.

Many of the observed masing sites have individual masers located along
lines or arcs, often with a near-monotonic velocity gradient along the
line. It was suggested \citep{norris_93, norris_98, phillips_98} that
these linear structures trace masers embedded in an edge-on disk
surrounding a young star. Values of disk radii and enclosed masses,
derived from modelling these sources assuming Keplerian rotation,
agree with theoretical models of accretion disks around massive stars
\citep{lin_90, hollenbach_94}.
Recently a subsection of one methanol maser has been extensively and
successfully modelled as a disk \citep{pesta_04} amplifying a
background source, but on the whole these models do not satisfactorily
explain the majority of methanol masers.

Whilst Norris \etal\ claimed the observations demonstrated they
delineated disks, \cite{walsh_98} found only 36 of 97 maser sites to
be linearly extended. While the circumstellar disk hypothesis is
consistent with their data, they considered it unlikely as the derived
values of the enclosed masses are too high in some cases. They
suggested the linear geometry of 6.7-GHz methanol masers resulted from
shocks with a small, smooth, velocity gradient, as a consequence of
the distance of the masers from the shock origin. \cite{norris_98}
had rejected this as a possible explanation as they believed shocks
could not produce the velocity gradient they saw in many of their
maser sites. 
%

Edge-on shock and edge-on disk models produce linear features across
the sky but shock fronts are not normally so well ordered. At mas
resolutions we probe structures of the order of tens of AU, the
perfect scale to investigate these differences. One further crucial
difference is that the first model would have the magnetic field
across the major axis of the feature -- as the disk is formed by
matter collapsing along the field lines \citep{Mo1999b}, and the
second will have it along the axis, due to entrainment of the field at
the shock front. Therefore measuring the field directions could
distinguish between these two models.

Detailed studies \citep{watson_94,elitzur_96c,gray_03} of the polarisation
properties of OH, \water\/ and SiO masers have been made, but there
has been relatively little investigation of the polarisation
properties of methanol masers.  Single dish observations of
polarisation in methanol masers are all from the strongest sources:
\cite{Ko1988} observed W3(OH) and NGC~6334F at 12~GHz, while
\cite{Ca1995a} determined that the level of circular polarisation for
a number of strong 6.7~GHz methanol maser features was less than
1\%. \cite{ellingsen_02} reported on 6.7~GHz polarisation observations
with the ATCA of NGC~6334F, where he found linear polarisation
fractions of up to 10\%. The first high resolution polarisation images
were made with MERLIN at a resolution of 50~mas \citep{vlemmings_06},
of W3(OH), where polarisation fractions of upto 8\% but typically
2-3\% were found.  Those polarisation angles were consistent with
those of the OH-masers, and furthermore lie along the methanol maser
emission as expected for shock excited regions.

\subsection{G339.88-1.26}

The maser G339.88-1.26 was discovered at 12~GHz by \cite{norris_87},
and at 6.7~GHz by \cite{macleod_92}.
%
Follow-up at arcsecond resolution by the Australia Telescope Compact
Array (ATCA) showed strong emission with a roughly linear morphology and a
monotonic velocity gradient \citep{norris_93}. 
\cite{ellingsen_96_cont} detected weak radio continuum emission which
peaks at the same position as the maser emission.  The Mid Infra-Red
(MIR) observations at $10~\mu$m of \cite{St1998a} detected elongated
emission along the same position angle as the disk inferred by the
methanol masers. These combined to make this one of the best
candidates for the disk model for methanol masers, with the masers
formed in the dusty disk hiding an embedded massive, forming, star.
However higher resolution, 10 and $18~\mu$m MIR observations, at the
Keck Observatory \citep{debuizer_02}
%
resolved the IR source into three components. VLBI observations from
1996, included in the same paper, showed that the masers formed an
inverted Y shape and lay between 1B and 1C. See Figure 2b in that
paper. That is they do not mark the disk, indeed there was no disk,
nor do they have a Keplerian velocity distribution.
%

\section{Development of new Mount types in AIPS} 

As a telescope tracks a source across the sky the angle of the
source on the sky to that of the telescope feed changes. When
observing with Left Circular and Right Circular Polarisations (LCP and
RCP) this has the effect of introducing a phase shift between the two
recorded data-streams. To post-process these data these phases need to
be removed, and the feed impurities need to be solved for and included
in the calibration. A good summary of the steps needed to make
polarisation VLBI images can be found in \cite{aaron_97}. In this
study, and indeed all other sources on VLBI polarisation, the feed
mounts are assumed to be Cassegrain or Equatorial. Different mounts
rotate the feeds in different fashions on the sky as the telescope
tracks a source. Two new mount types to those supported have been
added.

For VLBI observations AIPS \citep{aips} remains the only tool for
data calibration, therefore this has been our target for the extension
of mount types. The code to support the Nasmyth mount type and the E-W
mount type have been developed. The latter is a subset of the X-Y
mount (traditionally for Low Earth Orbit satellite tracking stations)
where the second axis lies East-West. The only example known to the
author is the Hobart telescope which is part of the Australian Long
Baseline Array (LBA). The alternative configuration is the N-S mount,
where the second axis lies North-South. The `keyhole', where large
angular changes are required for small movements on the sky, falls on
the second axis. Compared the Alt-Az mount, the X-Y mounts move the
keyhole from the Zenith to the horizon, where observations are not
normally made. The second mount type added, which is not used in the
data presented here, is for the Nasmyth type. The Nasmyth mount is
more normally used on optical instruments, as it allows space for very
large instrumentation packages. Until now this has not been needed for
Radio Telescopes, but new quasi-optical systems allow the siting of
multiple feeds for different bands at the Nasmyth foci. This
configuration allows the co-observing of widely separated observing
bands, which is particularly useful for the calibration of mm-VLBI
\citep{fpt_report}. The new code is for the new mm-VLBI
telescope being constructed at Yebes, Spain, which will cover
frequencies from 2\,GHz to 115\,GHz, and also the IRAM telescope at
Pico Veleta (Spain). Left handed and Right handed Nasmyth foci are
also included in the new AIPS code, as required for the Left or Right
optical branch. Full details can be found in \cite{pol_report}. These
observations with the LBA were used to test EW-mount portion of these
new subroutines.

\begin{figure}
\begin{center}
 \epsfig{file=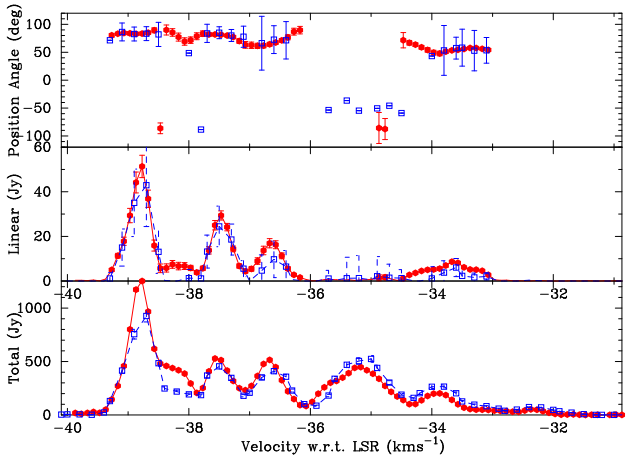,width=6cm,angle=270}
 \caption{Polarisation angle, linear and total flux for G339-1.26, as
   observed by the LBA (this paper, blue open squares) and the ATCA
   (Ellingsen, priv. comm., red closed circles). The spectra is scalar
   summed across the image (Stokes I, Q and U), and shows good
   agreement between the VLBI and the connected array results. The
   found Stokes I flux from the 2001 observations is slightly higher
   between ($-34.5$ to $-33.0$ \kms ), whereas the fractional
   polarisation is slightly lower.  This is the region found to be
   varible in \cite{goe_04}.  The errors are the absolute errors based
   on the confidence in the polarisation calibration (2\% and 0.4\%
   respectively), not the relative errors. Where errors are not shown
   they could not be calculated.}
\label{fig:pa}
\end{center}
\end{figure}

When the LBA results on G339.88-1.26 are compared to those of ATCA
observations (Ellingsen, Priv. Comms.), made in Sept
1999, an excellent match in the linear polarised fluxes and angles
is found, after
including the absolute polarisation angle offset between the brightest
polarised component (at $-38.7$\,\kms ). See Figure \ref{fig:pa} where
the position angles, linear polarised and total flux are compared for
the two instruments. ATCA values are plotted with red closed circles,
and the LBA values with blue open boxes. The values are extracted from
a sum across the Q and U images (with miriad's {\bf imspec}) which
allows the comparison of these two datasets with very different
resolutions.  The errors are absolute errors from the confidence in
the polarisation calibration, the relative errors are much less. Note
that \cite{goe_04} report this source as `not significantly variable',
with one component falling (-32.39 \kms ) and one rising (-33.19 \kms
).  Both of these appear in the Western cluster, which is where we
find great emission compared to 1999. There is also a short-fall in
flux (total and linear) from between -38.5 and -38 \kms .

\section{Observations}
\label{s:obsdata}

Observations for this experiment, V148, were performed in two
sessions, V148A in 25 April and V148B in 23 July 2001. Observations
were made of the calibrators; 1718-649 and 1610-771 plus 1921-293 and
1334-127 for fringe and absolute position angle calibration and three
target masers; NGC~6334F, G327.120+0.511, G339.88-1.26. Unfortunately
in the first set of observations Hobart failed, leaving only four
antennae; Parkes, Mopra, ATCA, and Ceduna. These were not sufficient
to provide good polarisation calibration. Only total power
observations could be extracted from this session. The V148B
observations included Hobart and could be calibrated. Of these
observations only G339.88-1.26 produced
interesting results, so this paper only reports on that source.
The other two targets were effected by a range of correlator issues,
and will be reobserved. 
The data recorded were 4~MHz of 2 bits, dual polarisation, and all
four stokes and 1024 channels were formed at the LBA correlator. The
data sampling was two seconds, and the autocorrelations were also
recorded to aid calibration with the AIPS task ACFIT. A problem with
the sampler levels, which leads to ringing across the bandpass, meant
that the data had to be Hanning smoothed over adjacent channels,
giving an effective velocity range of 0.35\,\kms .
The resolution of these observations is $3.1\times2.8$ mas, with a
position angle of $49^o$. The RMS for a single channel, unaffected by
residual flux, is 0.1\,Jy.

The calibration was done via the normal routes as described in
\cite{diamond_spec_vlbi}, and as reported in \cite{dodson_04}.  Delay
and rate calibration was done in AIPS, and then final self-calibration
and model-fitting in difmap \citep{difmap}. In addition polarisation
imaging was done in Miriad \citep{sault_95}, on the
difmap self-calibrated data. AIPS tasks are refered to in uppercase bold,
and Miriad tasks in lowercase bold. 
1718-649, being a point-like source, provided the final amplitude
calibration as the flux (2.7~Jy) was known from ATCA observations. The
polarisation corrections were made with the total intensity models of
the sources 1718-649 and 1610-771 derived from both epochs, using the
code with new mount types defined, and the task {\bf LPCAL}.
As this is the first demonstration of the code great care has been
taken to ensure that the phase corrections produce the expected
behaviour, on this experiment and also on others available
(V182). See \cite{pol_report} for details of the analysis. The
polarisation corrections, the D-terms, were found to be very high -- as high
as 20\% for some antennae -- which could perhaps be expected on this
very early full-polarisation VLBI experiment for the LBA. Recent
results (from the e-VLBI real time monitoring) indicate that the
polarisation purity has improved greatly.
Nevertheless, the estimated accuracy of our polarisation calibration
is about 2\%, based on the RMS difference between solutions on the two
different calibrators -- both assuming that the sources were polarised
and unpolarised.
The data cubes were formed at full spectral resolution (even though
the data had been Hanning smoothed) at 0.5 mas/pixel in difmap. All
channels were searched for emission across the source, and simple
Gaussian models were fitted where emission was found. The data was
self-calibrated on the strongest compact emission (at -37.5\,\kms ) to
the model. The data were exported to Miriad, imaged at 0.7 mas/pixel
and cleaned around the regions where the difmap models were found.
The regions in which no emission had been found were blanked.
Deconvolved images of Stokes I, Q and U were generated and
polarisation images were derived from these, clipped at a position
angle error of $5^o$, using {\bf impol}. The polarisation images
presented are from the images integrated over frequency, as the spots
are isolated in space and velocity. This allowed more compact datasets
to be formed without loss of information. There is little noticeable
difference between the images from the entire datacube and the
integrated one. 

\section{Stokes I images of G339.88-1.2}

\begin{figure}
\begin{center}
\epsfig{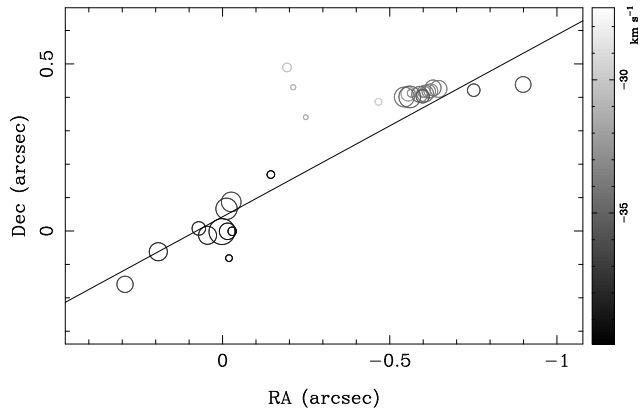}
\caption{Spot plot of G339.88-1.2. The sizes represent the
square root of the component flux and the colour the velocity. There
are three clusters of points, to the North, the East and the West. The
position errors are much less than the symbol sizes. The line shows
the best fit to the disk, at $-61^o$.}
\label{fig:pos}
\end{center}
\end{figure}

\begin{figure}
\begin{center}
\epsfig{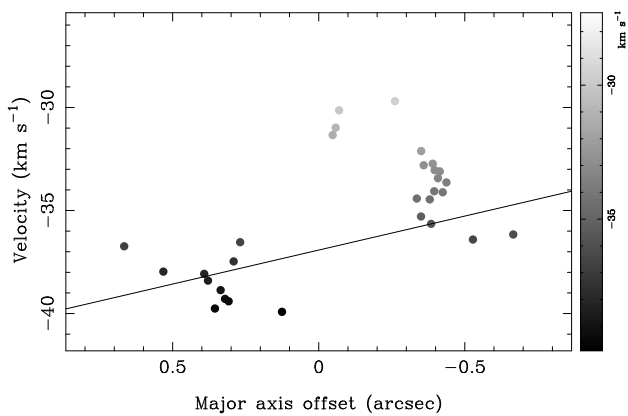}
\caption{Velocity-major axis plot, showing the distance along the
  major axis (the best fit to the disk) against velocity. The spots
  are extended in velocity perpendicular to the major axis, and the
  fit as a linear distribution due to emission from a ring around a
  central gravitational source -- which gives 11\,\msol\ (shown) -- is
  very poor.}
\label{fig:vma}
\end{center}
\end{figure}

We present Stokes I images of the source in Figure \ref{fig:pos} and
\ref{fig:vma}, prepared using {\bf
msplot}\footnote{http://www.atnf.csiro.au/people/Chris.Phillips/software.html}.
Figure \ref{fig:pos} is a `spot plot' derived from the models fitted
to the uv-data in Difmap. This allows us to present a noise free
image, with all the components clearly identified. The velocities of
the components are colour coded, whilst the square root of the
flux/beam sets the circle size. This provides both a comparison to
other, similar, papers (e.g. \citet{dodson_04}) and a guide to the
arrangement of the different features, whilst not suppressing the
weaker features. The map is self calibrated to the strong compact
feature at $-37.5$ \kms , and this then falls at the origin.
We find the lowest absolute velocity, and weakest, spot positions to the
North. These run from -28 to -31.5\,\kms , with most negitive velocities
furthest from the centre. This is also the brightest feature peaking
at at 9~Jy/beam.
The Western cluster starts at -32.5\,\kms , and runs further westwards
with velocity until -33.5\,\kms . After which the spots track back
across the same positions in the sky. There is some further emission
in the extreme West around $-36.5$\,\kms . The Eastern cluster appear
at about $-36.5$\,\kms\ and run in a easterly direction until around
$-38$\,\kms , after which the components appear to the South of the
starting point.

Figure \ref{fig:vma} plots the velocities of the features as a
function of their distance along the major axis (shown as the line in
Figure \ref{fig:pos}) from the centre of the emission along this line,
the velocities are along the y-axis. Emission from a thin ring in a
disk would produce a linear distribution.
The detailed mapping of spot position against velocity is difficult to
reconcile with that expected from an edge-on Keplerian rotating disk
model, as there is not a simple two sided structure let alone a
monotonic increase in velocity along the main axis. But other models
are as difficult to construct.
If the emission delineates an entire disk, gravitationally bound by a
massive central object, at a distance of 3\,kpc, the implied enclosed
mass is 11\,\msol .  However, in this source the spots are extended
across the major axis, as discussed in \cite{dodson_04}, and
inconsistent with the disk-model. Similar structures are seen in the
observations of water maser by \cite{torrelles_02}, on both the small
and the large scale. Water masers unquestionably form in shocked
out-flows. These facts combine to argue strongly against the
disk-based models for the majority of the emission.


\section{Measuring the polarisation of 6.7~GHz methanol maser spots}

From the polarised images we present fluxes and polarisation angles
for all the spots, defined as regions of contiguous emission in space
and velocity, in tabular form. Figures are also presented, showing the
arrangement of the emission and the magnetic fields that this
represents. The Northern region, being very weak, has no identifible
polarised flux. The Western has rapidly varying linear polarisation
angles ($\chi$) which tend to follow the ridge of
emission. Interestingly the direction undergoes a swap across the
brightest point, changing by $90^o$ going North to South.  Across the
North-Eastern region, at the phase centre, $\chi$ is similar. In the
most Eastern region $\chi$ is again variable. In the South-Eastern
region $\chi$ the flux emitting region is continuous and $\chi$ is
constant across this.

To relate $\chi$ to magnetic field direction is not simple.  The
dominant maser emission will either be parallel or perpendicular to
the magnetic field lines, depending on the angle between the magnetic
field and line of sight. If it is greater (less) than $\sim 55^o$ the
vectors are perpendicular (parallel) to the magnetic field
\citep{gold_73}. Also when the masers are highly saturated there are
further complications, but as the polarisation vectors are similar
from all components, weak and strong, one can deduce that the emission
is never sufficiently saturated to change the emitting
regime. Furthermore, the source is extended on the sky, being over
1\arcs\ across. It is unlikely, therefore, that the field is pointing
towards us.  The one region where this may not be correct is around
the brightest spot in the Western cluster, where the polarisation
vectors rapidly change by $90^o$ across it, yet the polarisation is
zero at the centre. Very similar behaviour is also seen in the
strongest feature reported on the \water\ masers discussed in
\cite{vlem_06}. For our source better data is needed, in particular
the magnetic field magnitudes, to solve for the full field behaviour
using Stokes V. (As Methanol and OH masers are often found co-located
(as in W3(OH) \citet{vlemmings_06}) we could expect the field
strengths of $\sim -4$ to $-6$\,mG as found in \citet{caswell_03} and
\citet{caswell_04} for OH masers at 6-GHz and 1720-MHz respectivily.)
Therefore we have assumed that we are in the regime where the
polarisation vectors are perpendicular with the fields, and in all
figures the polarisation vectors have been rotated by $90^o$, to show
the magnetic field directions.
Models for the 2D galactic magnetic field have recently been derived
\citep{brown_07}, so a line of sight rotation measure (RM) ($-7 \pm 3$)
can be deduced and used to solve for the magnetic fields.
This low value for the RM is because the maser sits near a reversal in
the galactic magnetic field, and all contributions cancel. The nearby
pulsars J1625-4048 and J1703-4851 have RMs of $-7\pm15$ and
$-4\pm24$. The model RM implies an angle change of +0.8 degrees at
6.7~GHz. This is less than our estimated errors (which are less
than $5^o$ for all but the edges of the images), and so it has not
been included.

\begin{figure}
\begin{center}
\epsfig{file=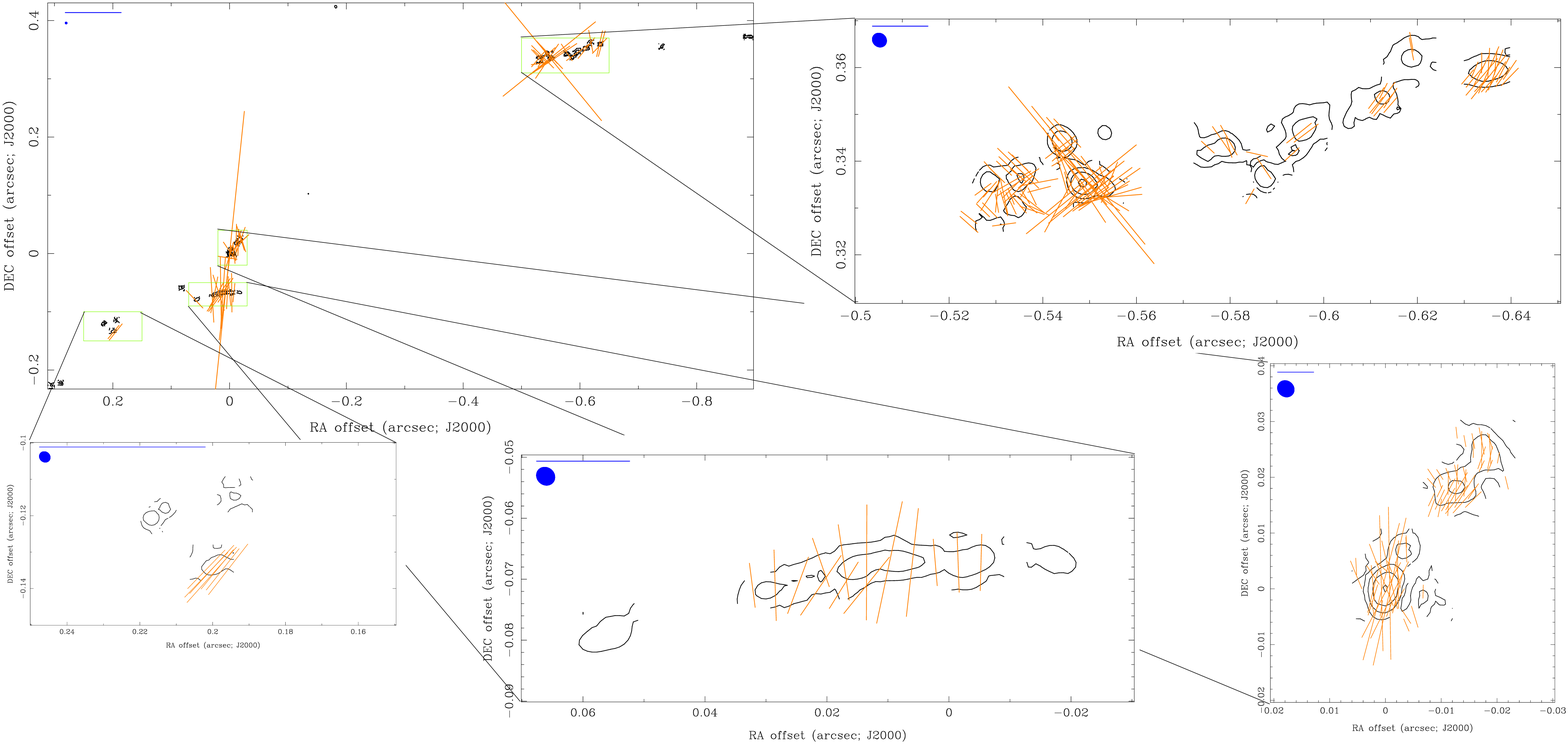, width=25cm, angle=270}
\caption{Total flux (contours at 2,8, 32 and 128 Jy/beam) overlaid
  with the polarisation intensity and direction (scaled vectors). The
  beam size shown in top left, along with bar representing 10~Jy of
  linearly polarised flux. The polarisation vectors are rotated to
  represent the magnetic field. The four regions with significant
  polarised emission are expanded.}
\label{fig:pol}
\end{center}
\end{figure}

Figure \ref{fig:pol} shows the total flux in contours, with the
vectors representing the magnetic fields, 
without the small correction for RM, scaled by the linear
polarised flux intensity. The regions of interest are expanded. Figure
\ref{fig:se} shows the South Eastern component from Figure
\ref{fig:pol} but with the total flux as an image, the linear polarised in
contours, and the vectors representing the magnetic field overlaid
scaled with the fractional polarisation.

The major axis of the cluster lies at $-61^o$ (all angles are measured
from north through east), and the majority of the vectors follow the
ridge of emission, except for the reversal in the Western cluster, and
the points to the south of the phase centre.
The magnetic fields are expected to lie along the flow of material if
the material is in a shock. For a disk the magnetic fields would be
expected to thread through the disk and be at right angles to the
major axis, which is clearly not the case across the majority of the
emission.
However, we note with interest that the region with the most
continuous surface brightness, the Southern bar of the Eastern
cluster, from a single narrow spectral feature spanning -38 to
-39.5\,\kms , has magnetic fields which lie as predicted by the disk
model, that is perpendicular to the emission. See Figure \ref{fig:se}a in
which Stokes I, overlaid with linear polarisation flux contours, and
the vectors (rotated to represent the magnetic field) of the
fractional polarisation and direction is plotted.
The velocity-major axis plot (Figure \ref{fig:se}b) shows a linear
relationship, but implies only a extremely small contained mass of
0.03\,\msol . 

In Table 1
the data is summed (in I, Q and U) over each `spot' in the full data
cube, after masking for greater than 2~Jy/beam in stokes I, and the
Stokes I, the linear (in janskys) and fractional (in percent)
polarised flux measurement is derived. Furthermore a Guassian was
fitted to the summed channels and the found extent (major and minor
axes in mas and the position angle in degrees) is reported. This does
not (except is the few cases where the emission is from a single
component) represent the spot sizes, but the region over which
emission is to be found. Finally the centre of the region and the
velocity range (in mas for $\Delta \alpha$ and $\Delta \delta$, and
\kms for the velocity) is given.

\section{Discussion}
\label{s:discuss}

The debate for and against methanol masers forming in disks still
rages a decade after it began. Higher resolution observations, IR and
VLBI, are slowly throwing light onto the exact conditions of these
sources. Certainly shock based models seem to be the only ones which
can reproduce more than the most localised or gross features of
methanol masers in massive star-forming regions.
Even the ring of masers discovered by \cite{ring} do not
fit neatly with the expectations of maser in a disk, despite the fact
one of the major arguments against the disk model was the lack of top
view masers. These masers also show a similar extension perpendicular to the
major axis, in this case the ring around the central source. 
Nevertheless, the analysis of \cite{pesta_04}, of one component of the
maser emission in NGC~7538, indicates that probably disks do host some
masers. Possibly a similar region is found in G339.88-1.26, in
which the magnetic field lies parallel to the emission. However, in
both cases, the majority of the source cannot be fitted in this
fashion.

The extent of the SE bar (Figure \ref{fig:se}), given the distance of
3~kpc based on the systematic velocity, is 200~AU in size and only
1.5~\kms\ in velocity. This leads to a contained mass of 0.03\,\msol .
This disk size is similar to what has been found in other star-forming
regions, and if the region is made up of sub-massive stellar bodies,
which will combine to form the eventual massive star, such an object
is possibly feasible. However such an object could only survive where
it was not disturbed by external turbulence, which is unlikely in a
massive star-forming region.
\cite{debuizer_02} places the methanol masers on the edge of
MIR source 1B, and suggest that they are formed by material flowing
from that and interacting with 1C. They suggest that an undetected (in
MIR) foreground object is responsible for the radio and visible
detections and estimate this to be of spectral type B2.5.
One could assume, therefore, that this emission is from the
front edge of a disk around such a star, as in \cite{pesta_04}. The
similarity of this object to the disk feature in NGC~7538 is striking,
but as there are no good 12~GHz images to compare with, the complete
analysis performed by \cite{pesta_04} is not immediately possible.

The review of \cite{vlemmings_07} draws attention to
the conclusions which can be drawn from the Stokes-V analysis of VLBI
images. No attempt has been made to derive Stokes-V from the data, as we must firstly improve the polarisation purity. 
These are the first results in linear
polarisation from the LBA, the polarisation corrections found are
high, and it would be more sensible to use these results as a
stepping stone to such interesting analysis.

\section{Conclusions}
\label{s:conclude}

The first polarisation results from the LBA, and the first VLBI
polarisation images of methanol masers, are presented. 
The polarisation calibration of the LBA is undergoing
development. This experiment, we believe, provides a first successful
demonstration of the process. However, further work, on multiple
targets, is required to provide complete confidence in the
system. These observations will demonstrate the successful application
of the calibration to a range of sources in the sky. Once this is
complete the code will be incorporated in the AIPS distribution.

\begin{figure}
\begin{center}
\epsfig{file=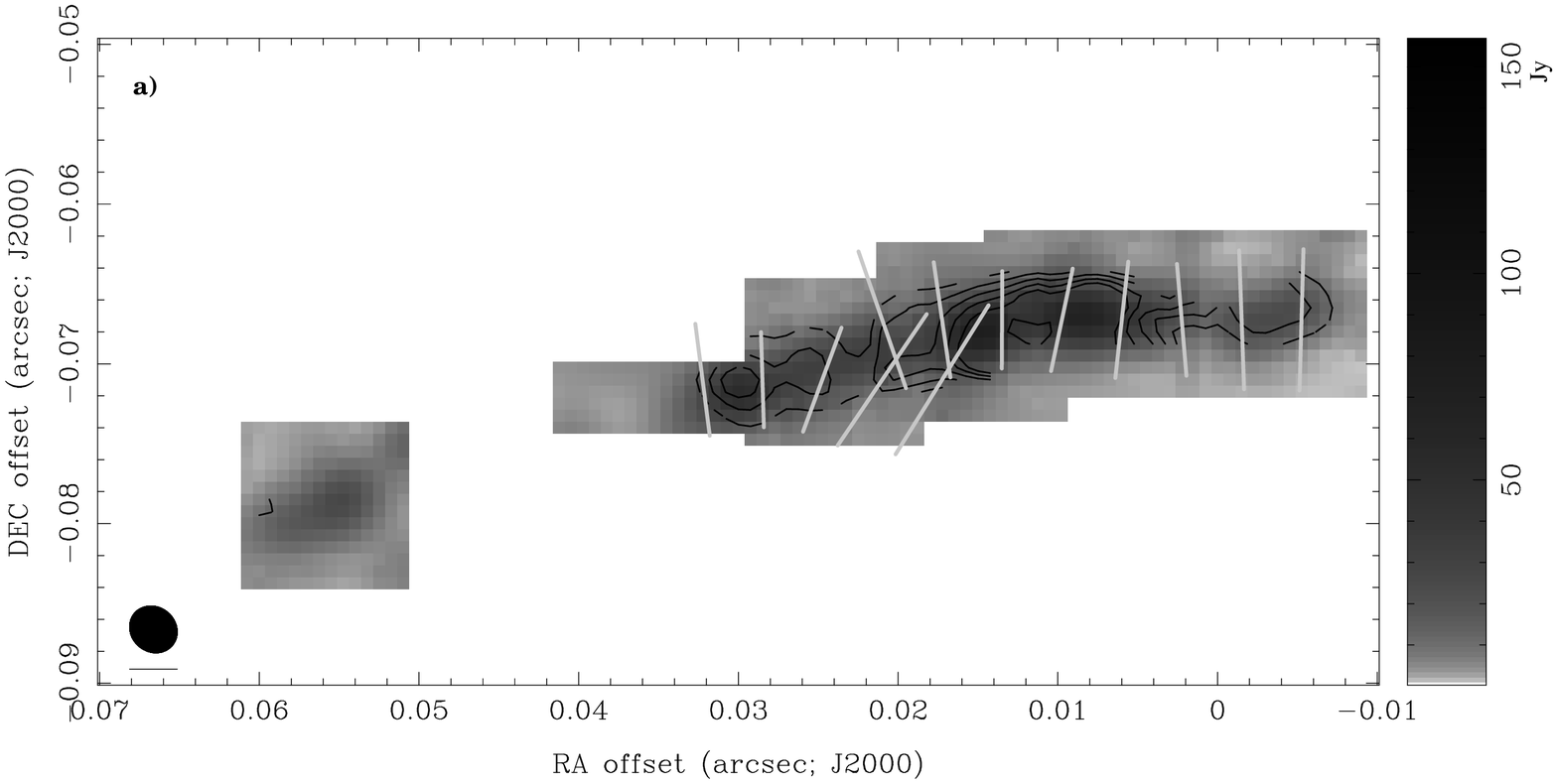, width=8cm, angle=0}
\epsfig{file=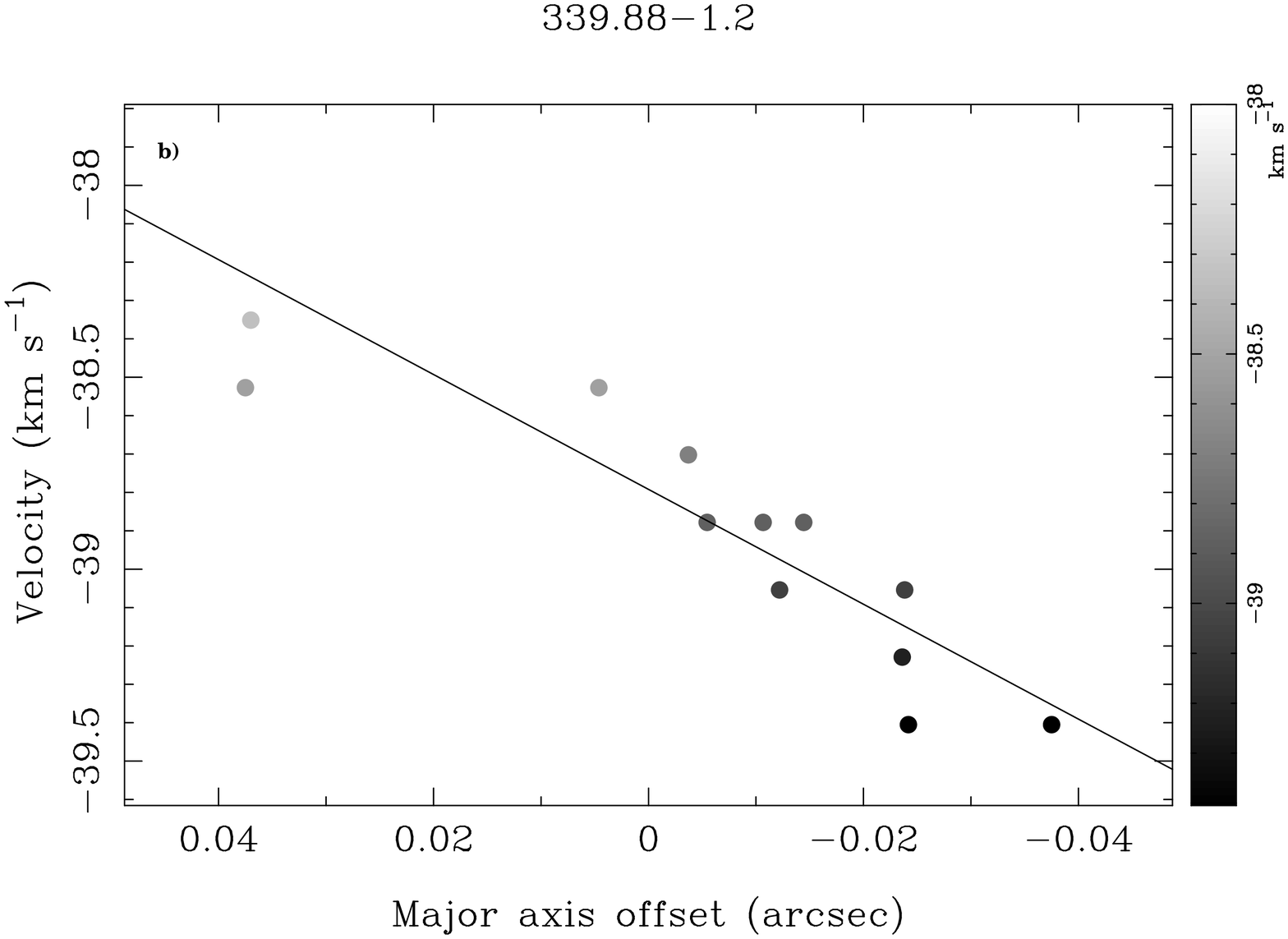, width=8cm, angle=0}
\caption{a) Image of the SE feature with magnetic fields lying
perpendicular to it. The image shows the total flux image,
along with a colour bar, in Jy/beam. The linear polarised flux
contours (4,6,8,10 and 12 Jy/beam), and the polarisation vectors,
scaled to the fractional polarised flux, rotated to represent the
magnetic field direction. The beam size and a bar representing 10\%
fractional polarised flux is shown in the bottom left.
b) The Velocity-Major Axis plot shows that this feature is linear in
velocity space, and therefore could be represented by a ring (or
partial ring) of emission around a central gravitational source.}
\label{fig:se}
\end{center}
\end{figure}

The target of our observation shows a disordered magnetic field, which
in most regions lies along the ridge of emission. This is in
accordance with shock generated features. However there is one small region
in the complex which conforms to the expectations of a disk. The
Position and Velocity-Major Axis plots (Figure \ref{fig:se}) show
linearity in space and velocity and the magnetic field lies parallel
to the feature. However it cannot represent a complete disk, as the
deduced enclosed mass is extremely small (0.03\,\msol ). Furthermore the
narrow waist of NGC~7538 and evidence of an out-flow at right angles
to the disk is missing in this case.
Further observations, with improved polarisation calibration, and also
at 12.2~GHz, are required to investigate this. 

\acknowledgements I wish to thank Dr Brown for calculating the RM
for the direction and distance of this source, and Dr Ellingsen for
making the ATCA polarisation data available. The referee, Dr Vlemmings, 
made comments which greatly improved the paper. 
This research has made use of NASA's Astrophysics Data System Abstract 
Service and the SIMBAD database, operated at CDS. 
The Long Baseline Array is part of the Australia Telescope which is
funded by the Commonwealth of Australia for operation as a National
Facility, managed by CSIRO and the University of Tasmania.
I acknowledge support by a Marie Curie International Incoming
Fellowship within the EU FP6 under contract number
MIF1-CT-2005-021873. The Raman Research Institute provided hosting and
support during the completion of this paper.

\onecolumn

\twocolumn  

\onecolumn
\begin{table}
\begin{scriptsize}
\begin{center}
\begin{tabular}{rrrr|rrr|rrrc}
\hline
\multicolumn{4}{c}{Integrated Flux}&\multicolumn{3}{c}{Region Size}&\multicolumn{4}{c}{Position and Velocity}\\
I & linear & fraction & $\chi$ & Major & Minor & PA & $\Delta \alpha$ & $\Delta \delta$ & Peak & Range \\
Jy & Jy & \% & $^o$ & mas & mas & $^o$ & mas & mas & \kms & \kms\\
   3.3 &  -   &    - &   - &  4.0 &  2.0 &   15 & -182 & 424 & -30.1 &  -29.9 -- -30.8 \\ 
\hline
  43.3 &  -   &    - &   - &  1.6 &  0.9 &    8 & -544 & 344 & -32.4 &  -31.9 -- -32.9 \\ 
  22.2 &  2.2 & 10.1 &  57 & 13.7 &  5.6 &  -66 & -612 & 353 & -33.3 &  -32.8 -- -33.6 \\ 
  54.2 &  3.7 &  6.8 &  62 &  6.6 &  2.3 &  -82 & -635 & 359 & -33.6 &  -33.3 -- -34.2 \\ 
  43.5 &  1.5 &  3.5 &  75 &  7.9 &  4.5 &  -60 & -619 & 362 & -34.0 &  -33.6 -- -34.5 \\ 
  44.6 &  1.7 &  3.8 & -63 &  7.8 &  2.4 &  -76 & -577 & 342 & -34.5 &  -34.2 -- -34.9 \\ 
 135.7 &  2.0 &  1.5 & -32 & 12.6 &  8.5 &  -46 & -534 & 335 & -34.9 &  -34.5 -- -35.6 \\ 
 104.6 &    - &    - &   - & 52.0 &  9.2 &  -23 & -558 & 392 & -34.9 &  -34.9 -- -35.6 \\ 
 244.7 &  1.2 &  0.5 & -71 &  2.0 &  1.1 &   87 & -549 & 335 & -35.4 &  -34.7 -- -35.7 \\ 
  36.2 &  0.9 &  2.5 & -44 &  2.2 &  1.5 &  -48 & -588 & 337 & -35.7 &  -35.4 -- -35.9 \\ 
  59.0 &  1.8 &  3.0 & -81 & 26.3 &  1.5 &  -88 & -887 & 372 & -36.3 &  -35.9 -- -36.4 \\ 
  20.2 &  0.6 &  3.1 &  27 &  3.4 &  1.5 &  -28 & -740 & 355 & -36.4 &  -36.1 -- -36.8 \\ 
\hline																	   
 148.0 &  6.9 &  4.7 &  80 & 14.8 &  4.0 &  -43 & -14 &  20  & -36.6 &  -36.1 -- -37.1 \\ 
 197.9 & 12.5 &  6.3 &  86 &  3.4 &  0.6 &  -18 &   0 &   0  & -37.5 &  -37.0 -- -38.0 \\ 
\hline						     											   
  39.5 &  1.8 &  4.6 &  29 & 10.2 &  5.3 &  -61 & 306 & -227 & -36.8 &  -36.4 -- -37.0 \\ 
  71.2 &  1.9 &  2.6 &  59 &  8.6 &  4.4 &  -84 & 199 & -133 & -38.0 &  -37.7 -- -38.4 \\ 
\hline						     											   
  78.2 &  1.5 &  1.9 & -46 &  7.9 &  3.7 &  -61 &  56 & -79  & -38.4 &  -38.2 -- -38.7 \\ 
 375.7 & 21.4 &  5.7 &  83 & 23.9 &  3.1 &  -78 &  14 & -68  & -38.7 &  -38.4 -- -39.3 \\ 
  55.8 &  4.0 &  7.2 & -76 &  8.4 &  3.2 &  -76 &  -3 & -67  & -39.1 &  -38.9 -- -39.6 \\ 
\end{tabular}
\end{center}
\label{tab:fluxes}
\caption{Physical parameters for the emission in G339.88-1.2. Each region
  of emission, masked below 2 ~Jy/beam of total flux, is
  averaged over the channels and the integrated flux (Stokes I, and
  linear) was found. These are given in janskys (i.e. mean power in
  frequency for the region), and in percent. Where the flux found was
  less than 0.5~Jy (4 $\sigma$ for the linear polarised flux) no value
  is given.  The size of the region (from a Gaussian fitted to the
  summed data) is given in mas and degrees. The
  maximum of this region is given in mas and \kms . The range of
  velocities over which the average flux was taken is given in \kms . }
\end{scriptsize}
\end{table}

\end{document}